\documentclass[aps, prl, twocolumn, superscriptaddress, nofootinbib]{revtex4}
\usepackage{amsmath,amssymb,amsthm,xspace,mathtools,graphicx,paralist,braket,hyperref}%
\usepackage[T1]{fontenc} 
\usepackage[usenames,dvipsnames,table]{xcolor}
\usepackage{soul}
\usepackage{enumitem}
\hypersetup{
    colorlinks=true,    %
    urlcolor=blue,      %
    linkcolor=blue,    %
    citecolor=blue      %
}

\definecolor{darkblue}{RGB}{0,0,158}
\definecolor{lightgrey}{RGB}{220,220,220}

\newtheorem{theorem}{Theorem}
\theoremstyle{definition}
\newtheorem{definition}{Definition}

\usepackage{color,graphicx}
\newcommand{\look}[1]{\textcolor{blue}{[Comment: #1]}} %

\newcommand{\psim}{\ket{\psi}_{AB},\{M_a^x\},\{M_b^y\}}
\newcommand{\spec}{\mathsf{spec}}

\long\def\old#1{}

\begin{document} 

\title{\vspace{-1cm}Impossibility of adversarial self-testing and secure sampling}

\author{Akshay Bansal$^*$} 
\affiliation{Department of Computer Science, Virginia Tech, Blacksburg, Virginia 24061, USA}

\author{Atul Singh Arora\footnote{The first two authors contributed equally.}} 
\affiliation{Department of Computing and Mathematical Sciences, California Institute of Technology, Pasadena, California 91125, USA
} 
\affiliation{Joint Center for Quantum Information and Computer Science (QuICS), University of Maryland \& NIST, College Park, Maryland 20742, USA}

\author{Thomas Van Himbeeck} 
\affiliation{Télécom Paris—LTCI, Inria, Institut Polytechnique de Paris, 91120 Palaiseau, France}

\author{Jamie Sikora} 
\affiliation{Department of Computer Science, Virginia Tech, Blacksburg, Virginia 24061, USA} 

\date{August 2024} 

\begin{abstract} 
Self-testing is the task where spatially separated Alice and Bob cooperate to deduce the inner workings of untrusted quantum devices by interacting with them in a classical manner. 
We examine the task above where Alice and Bob do not trust each other which we call \emph{adversarial self-testing}.  
We show that adversarial self-testing implies \emph{secure sampling}---a simpler task that we introduce where distrustful Alice and Bob wish to sample from a joint probability distribution with the guarantee that an honest party's marginal is not biased.
By extending impossibility results in two-party quantum cryptography, we give a simple proof that both of these tasks are impossible in all but trivial settings. 
\end{abstract}

\maketitle  
\paragraph{\textbf{Introduction.}}
The last few decades have witnessed massive leaps in the capabilities of quantum computers, in terms of  both theory and  implementation. With intensified efforts from government, industry and academia, the future where useful quantum computers are widely accessible, is becoming closer to a reality every day.
The availability of such quantum devices begs the question of whether one can trust that they are performing as advertised. 
In other words, should we blindly trust the output of quantum mechanical devices? 
And if not, is there a way to test them? 

Somewhat surprisingly, one can sometimes test spatially-separated devices to see if they are doing what they are purported to be doing based solely on its (classical) input/output behaviour and the assumption that quantum mechanics is a faithful description of Nature. 
This area is broadly referred to as \emph{self-testing}. 
As an early example, Coladangelo, Goh, and Scarani showed that any pure bipartite state 
can be self-tested~\cite{coladangelo2017all}. 
In addition, other strong results have been reported~\cite{mayers2004self, breiner2018parallel, coladangelo2017parallel, coladangelo2018generalization, bowles2018device, miller2013optimal, sikora2016minimum, bartusek2023secure}. 
Many of these deal with two parties, call them Alice and Bob, who \emph{cooperate} to ascertain the inner workings of the respective quantum devices.\footnote{Recent works~\cite{Metger2021, 2303.01293} give self-testing schemes of single devices using computational assumptions. We do not place any such limitations on Alice and Bob in this work.}

In this Letter, we prove that to self-test quantum devices, it is \emph{necessary} that Alice and Bob cooperate. %
More precisely, we define \emph{adversarial} self-testing as the self-testing task in the setting where Alice and Bob do not trust each other and give a surprisingly simple proof that this task cannot be realised. To this end, we use adversarial self-testing to perform a simpler task we call \emph{secure sampling}, and then leveraging results from quantum cryptography, show that secure sampling is impossible. We now introduce these two tasks in detail.

\medskip 
\paragraph{\textbf{Self-testing setting.}}
Consider a quantum mechanical device shared by two mutually trusting parties, Alice and Bob, each having their own part of the \emph{device} which we refer to as a \emph{box}. Each box has several buttons (input choices) and, upon pressing a button, one of several lights turns on (indicating an output). Crucially, suppose that the two boxes are not allowed to communicate after they are distributed to the parties (e.g. by ensuring enough physical separation between them).

The most general physical description of such a device is given by a \emph{device specification} $$\spec:=(\psim)$$ where $\ket{\psi}_{AB}$ is 
bipartite quantum state in an arbitrary Hilbert space, the projector $M^x_a$ corresponds to Alice inputting $x$ and obtaining outcome $a$, and similarly $M^y_b$ corresponds to Bob inputting $y$ and obtaining outcome $b$. We call the joint probability distribution of getting outcomes $(a,b)$ from the boxes, given inputs $(x,y)$, a \emph{quantum correlation} and denote it by 
\begin{equation}
  p(ab|xy)=\bra{\psi} M^x_a\otimes M^y_b \ket{\psi}. \label{eq:quantumCorrelation}
\end{equation}
Note that the marginals satisfy $p(a|x)=p(a|xy)$ and $p(b|y)=p(b|xy)$. %

\begin{figure}[h] \centering
  \includegraphics[width=4cm]{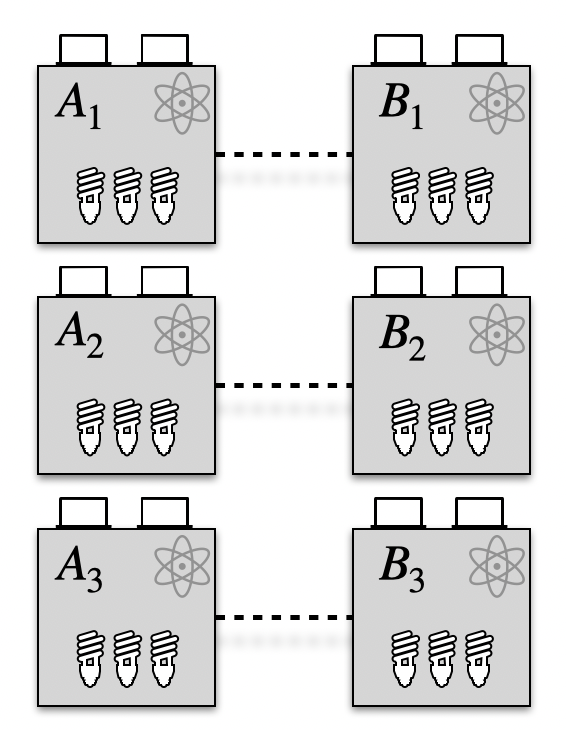}
  \caption{
  Alice and Bob share several pairs of boxes but have no reason to trust them. 
  They could test all but one and deduce (with hopefully high confidence) that either the boxes are faulty or the untested box will behave as expected. 
  } 
\label{Fig2} 
\end{figure}

In some cases, given that such a device produces a specific correlation $p(ab|xy)$, one can deduce the state $\ket{\psi}_{AB}$ and the measurements $\{M^x_a\},\{M^y_b\}$ up to local isometries, i.e. one can \emph{self-test} the device. In particular, one may be able to deduce that the state is entangled. In cryptographic contexts, self-testing allows one to model the quantum devices as black-boxes and thereby establish \emph{device-independent} security for tasks such as quantum key distribution. Treating quantum devices as black-boxes already includes the possibility that the quantum devices are prepared by the adversary and therefore yield secure constructions even against such powerful adversaries~\cite{acin2007device, pironio2009device, curty2014finite, vazirani2019fully}. %

How does one test whether the device, in fact, produces the purported correlation $p(ab|xy)$? Assuming Alice and Bob can communicate classically, they can start with $n$ devices and use $n-1$ randomly among them to \emph{test} for the right correlation~\cite{DIWCF}.\footnote{In the literature, the more common setting is where all $n$ devices are measured right away. See e.g.~\cite{vazirani2019fully} that shows how self-testing is used for key distribution. Note that, in both cases, the devices are \emph{not} assumed to be identical.}  More concretely (see also FIG.~\ref{Fig2}), Alice could randomly select $n-1$ devices, measure them and share this information with Bob who can then measure his part of the corresponding devices. He can then decide whether the purported correlation is consistent and share the result with Alice. If they are satisfied, they can use the remaining device, confident that it will work as claimed. %

The literature on self-testing and its use in quantum cryptography, in summary, points to the following recurring theme.

\vspace{0.5cm}
\textit{Theme: The ability to self-test suggests security in quantum cryptography.}  
\vspace{0.5cm}

Note that in the discussion above, even though Alice and Bob did not trust their quantum devices, they did trust each other. 

\paragraph{\textbf{Adversarial self-testing.}} Consider the fully-distrustful setting~\cite{PhysRevLett.106.220501} where Alice trusts neither her devices nor Bob, and similarly Bob trusts neither his devices nor Alice. Conceptually, define \emph{adversarial self-testing} to be the natural extension of \emph{self-testing} to this fully distrustful setting. 

We first look at why the self-testing procedure involving $n$ devices fails and then suggest a plausible alternative. %

In (standard) self-testing as described above, suppose Bob is malicious and he created the devices for his own nefarious purposes. If Alice suspects this might be the case, then she has no reason to trust the tests (that only he performed) nor his final decision to use the remaining device. 

To alleviate this issue, we allow the parties to exchange boxes---this can be achieved using quantum communication~\cite{PhysRevLett.106.220501}. We also allow them to prevent communication across the boxes they possess (e.g. by shielding). The idea is to have both parties perform tests and furthermore, if a party chooses to test device $i$, they ask for the corresponding box from the other party to run the test themselves using both boxes. %
A plausible protocol based on \emph{cut-and-choose} idea is for Bob to test roughly half the devices, selected uniformly at random, and then if he is satisfied, to allow Alice to perform her own tests. She can select all but one of the remaining devices and perform her own tests. If she is also satisfied, then they agree to use the remaining device. 

Why might such a strategy work? Since Alice and Bob's test involve randomly select devices, then regardless of who might have tampered with the devices, neither of them has full control over which device is ultimately used and as such, each device is likely to be tested.

\begin{figure}[ht] \centering
    \includegraphics[width=4.2cm]{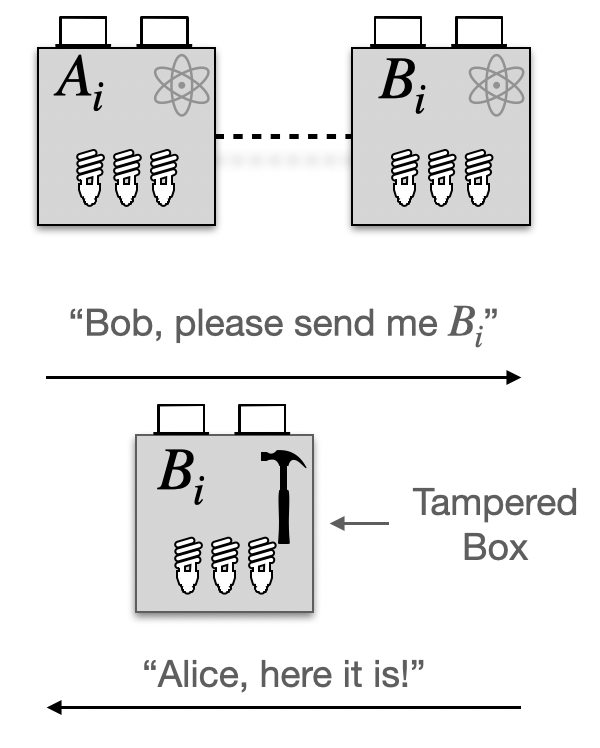}
    \caption{
If Bob is malicious, he can tamper with the boxes adaptively. 
For example, he can tamper with box $\Box_{A,i}$ when it was created, then change the contents of $\Box_{B,i}$ later before sending it to Alice.  
    } 
  \label{Fig3} 
\end{figure}

Before proceeding, we briefly remark on two important distinctions between the two settings. First, even though exchanging boxes is relevant for adversarial self-testing, as motivated above, it does not offer any advantage in the (standard) self-testing setting. This is because Alice and Bob can coordinate and broadcast their inputs and outputs and collectively process their statistics. %
Second, in (standard) self-testing, once the boxes are prepared and distributed, they can no longer be modified. However, for adversarial self-testing, a malicious party can tamper with the boxes at any point during the protocol as long as the box is in their possession. In fact, the only constraint is that the malicious party cannot tamper with the boxes \emph{currently} held by the honest party. For example, suppose Alice and Bob share two pairs of boxes and Alice asks Bob to send her his box from the first pair. %
Then Bob can tamper with his box right before he sends it such that it acts in a way favourable to him. See FIG.~\ref{Fig3} for an illustration.

What is known about adversarial self-testing? Recently, notions closely related to it have implicitly appeared in certain device-independent cryptographic settings, such as weak coin flipping~\cite{DIWCF} and network entanglement certification~\cite{PhysRevLett.131.140201}. 
In these works, only one-sided tests were used, i.e., where either Alice tests Bob or Bob tests Alice.
Moreover, for coin flipping, only partial security was obtained. %
This begs the question of whether one can have a (two-sided) adversarial self-test to achieve ideal security for such tasks? %
In this work, we show that this is impossible.

\paragraph{\textbf{Our contributions.}}

We start by concretely defining \emph{adversarial self-testing}. 
Consider a device specification $\spec:=(\ket{\psi}_{AB},\{M^x_a\},\{M^y_b\})$ and $n$ untrusted quantum devices purportedly consistent with $\spec$. %
Denote the $i$-th untrusted device by two boxes $\Box_{A,i}$ and $\Box_{B,i}$. Let Alice and Bob be two remote parties, connected by a classical channel and a quantum channel. Alice and Bob are uncorrelated initially except that Alice holds boxes $\{\Box_{A,i}\}_i$ and Bob holds $\{\Box_{B,i}\}_i$. Assume that, at any point during the protocol, boxes held by Alice cannot communicate with those held by Bob and also that the parties can choose to prevent communication among the boxes in their possession. %
Let $\cal{P}$ be a bipartite protocol among Alice and Bob, that specifies a procedure for the two parties to perform classical computations, exchange classical messages, measure and exchange $n-1$ of the $n$ untrusted quantum devices. Based on this, each party either outputs $\bot$ denoting abort or an index $j \in \{1,\dots n\}$ that specifies the \emph{certified} unused quantum device.

Consider the following three situations describing the various adversarial attacks we allow when the target specification is $\spec$. In each, an honest party proceeds exactly as specified by $\cal{P}$. %
\begin{enumerate}[leftmargin=*]
  \item \label{item:fullTrust} \emph{Fully trusted.} Both parties are assumed to be honest and all $n$ quantum devices are specified by $\spec$. %
  \item \label{item:untrustedDevice} \emph{Trusted parties (but untrusted devices).} %
  Both parties are assumed to be honest but all $n$ quantum devices are created by an adversary. %
  \item \label{item:fullydistrustful} \emph{Fully distrustful.} 
  An adversary creates the devices and controls \emph{one} party, say Bob, i.e. can make Bob arbitrarily deviate from protocol $\cal{P}$.
  The only restriction on the adversary is that it cannot influence Alice's classical computations and it cannot influence any quantum box while it is held by Alice. Similarly when Alice is controlled by the adversary. %
\end{enumerate}

We setup some notation.
We say that a device specification $(\ket{\phi}_{A'B'},\{N^x_a\},\{N^y_b\})$ is \emph{$\delta$-close} to the target specification $\spec$ %
if there are local isometries $\Phi_{A'}:A' \to AA''$ and $\Phi_{B'}:B' \to BB''$ such that 
\begin{align*} 
  \Phi (&(N^x_a \otimes N^y_b ) \ket{\phi}_{A'B'}) \approx_{\delta} \\ 
                      &(M^x_a \otimes M^y_b) \ket{\psi}_{AB} \otimes \ket{\mathsf{junk}}_{A''B''}
\end{align*}
for all $x,y,a,b$, where $\Phi:=\Phi_{A'}\otimes \Phi_{B'}$ and $\approx_{\delta}$ is used to denote that the states are at most $\delta$ far in trace distance. Then, adversarial self-testing is defined as follows.

\begin{definition}[Adversarial self-test] %
  $\cal{P}$ \emph{adversarially self-tests} the device specification $\spec:=(\psim)$, if there exist decreasing vanishing functions $\epsilon,\delta\ge 0$ such that the following conditions hold, corresponding to the three situations above:
  \begin{enumerate}[leftmargin=*]
    \item \emph{Correctness (Fully trusted).} %
    Both parties output the same device index $j$ (and neither aborts). %
    \item \emph{Self-testing (Trusted parties).} %
    Both parties have identical outputs and with probability at least $1-\epsilon(n)$, the protocol either aborts or, given that the protocol does not abort, the certified device $(\ket{\phi}_{A'B'},\{N^x_a\},\{N^y_b\})$ is \emph{$\delta(n)$-close} to the target specification $\spec$. %
    \item \emph{Adversarial self-testing (Fully distrustful).} %
    Suppose Alice is honest and Bob is controlled by the adversary. When Alice does not abort, denote by $\ket{\phi}_{A'B'}$ the purification of the state in Alice's certified box and assume its purification is held by Bob. Denote by $\{N^x_a\}$ the measurements corresponding to Alice's certified box. 
    Then, it is required that irrespective of Bob's output, the same condition as in the \emph{Self-testing} case above holds for \emph{some} measurements $\{N^y_b\}$.
    The analogous condition must also hold when Bob is honest and Alice is controlled by the adversary.
  \end{enumerate}
\end{definition}

We make two remarks about our definition. (i) Observe that if a specification $\spec$ %
can be adversarially self-tested then, in particular, $\spec$ can be self-tested in the standard setting (i.e. without exchanging boxes). %
(ii) A weakened variant of the third requirement, for instance, by assuming that one party, say Alice, is always honest, can be realised and has also found applications~\cite{PhysRevLett.131.140201, DIWCF}, as discussed earlier. %

Consider any device specification $\spec$ that produces a \emph{product correlation}, i.e. $p(ab|xy)=p(a|x)p(b|y)$ for all $a,b,x,y$ where $p(ab|xy)$ is as in Eq.~\ref{eq:quantumCorrelation}. This correlation can be produced \emph{locally} (without shared randomness) and \emph{classically}. %
Therefore, any guarantee from self-testing, which must be up to \emph{local} isometries, becomes meaningless. Remark (i) above entails that adversarial self-testing is, consequently, also meaningless in this case.
 
What can one say about device specifications that produce \emph{non-product correlations}, i.e. $p(ab|xy)\neq p(a|x)p(b|y)$ for some $a,b,x,y$? In \cite{coladangelo2017all}, the authors show that for every entangled state, one can find measurements such that the resulting device specification can be self-tested. Can one extend this to adversarial self-testing? The following result shows
that adversarial self-testing is impossible for \emph{any} meaningful specification. 

\begin{theorem} \label{thm1_}
  Any device specification $\spec$ that produces non-product correlations (i.e. $p(ab|xy) \neq p(a|x) \cdot p(b|y)$ for some $a,b,x,y$ where $p(ab|xy)$ is as in Eq.~\ref{eq:quantumCorrelation}) cannot be \emph{adversarially self-tested}.
\end{theorem} 

Adversarial self-testing of $\spec$ implies one can \emph{securely sample}, a simpler task we define below, according to the correlation $p(ab|xy)$ (which is produced by $\spec$ as in Eq.~(\ref{eq:quantumCorrelation})). Clearly, if secure sampling of $p(ab|xy)$ is impossible, the proof of Theorem~\ref{thm1_} is immediate. The remaining discussion focuses on proving the impossibility of secure sampling of non-product correlations. %

\medskip 
\paragraph{\textit{Secure Sampling.}}
We consider secure sampling in the bipartite setting where the parties are untrusted but their quantum devices are trusted. More concretely, a \emph{valid protocol} for \emph{securely sampling} from the joint distribution $p(ab|xy)$ is an interactive protocol among Alice and Bob who are given inputs $x,y$ and produce outputs $a,b$ (or abort). The protocol specifies local quantum computations for each party, involving an exchange quantum messages, to compute their respective outputs. 

As before, note that a malicious Alice may digress from the protocol. She may try to bias the marginal distribution of $b$. Similarly a malicious Bob may try to bias the marginal distribution of $a$. 
Note also that it does not make sense to consider the entire joint distribution $p(ab|xy)$ in the security analysis. This is because a malicious party can always output anything they wish, at the end. %
Thus, what makes sense is to bound the deviations away from the \emph{marginal distributions} $p(a|x)$ and $p(b|y)$. 

We say that $p(ab|xy)$ can be \emph{$\delta$-securely sampled} for a $\delta >0$, if there exists a protocol such that for any pair of outputs $(a,b)$, for any input $(x,y)$ the following hold.
\begin{itemize}[leftmargin=*]
  \item When both Alice and Bob are honest, they output $a$ and $b$ with probability $p(ab|xy)$.
  \item The probability that Alice outputs $a$ is at most $p(a|x) + \delta$ when Bob cheats (implying she did not abort). 
  \item The probability that Bob outputs $b$ is at most $p(b|y) + \delta$ when Alice cheats (implying he did not abort).
\end{itemize}
In words, a malicious party can only really influence the honest party's outcome towards  ``cheating detected''. We say that a distribution $p(ab|xy)$ can be \emph{securely sampled} if for any $\delta>0$, the distribution can be $\delta$-securely sampled. Note that, %
product correlations %
 can be trivially sampled securely---the parties sample their own outputs, depending on their respective inputs. We prove that the converse also holds. 
\begin{theorem} \label{thm2_}
  Given a \emph{correlation}\footnote{Recall that we only consider correlations/distributions that can arise from measuring a quantum state with local measurements.} $p(ab|xy)$, secure sampling is possible if and only if it is a product correlation. 
\end{theorem}  

From the definitions, one can check that adversarial self-testing of $\spec$ that produces $p(ab|xy)$ implies there is a protocol for secure sampling from $p(ab|xy)$. Thus, Theorem~\ref{thm1_} follows directly from Theorem~\ref{thm2_}. Turning to the proof of Theorem~\ref{thm2_}, we start with a simple fact about non-product correlations.

\medskip 

  \noindent \text{Lemma:}
  For a non-product correlation $p(ab|xy)$, there exist $a$, $b$, $x$, and $y$ such that 
  \begin{equation}
  p(ab|xy) > p(a|x) \cdot p(b|y). 
  \end{equation} 
  To prove this, suppose that for all $a$, $b$, $x$, and $y$ we have 
  \begin{equation}
  p(ab|xy) \leq p(a|x) \cdot p(b|y). 
  \end{equation}  
  Then, for any fixed $x$ and $y$, one can easily see that ${p(ab|xy) = p(a|x) \cdot p(b|y)}$ by adding over $a$ and $b$ on both sides and using the fact that if $0\le \Delta_i$ and $\sum_i \Delta_i = 0$, it follows that $\Delta_i=0$ for each $i$. 
  Thus, $p(ab|xy)$ is a product correlation, a contradiction.

\smallskip

The above lemma says that there is some input pair $(x',y')$ such that Alice and Bob's outcomes are correlated, i.e., not sampled from a product probability distribution. Henceforth, we focus on secure sampling of the non-product distribution $p(ab):=p(ab|x'y')$. %
This particular non-product distribution is exactly the issue when Alice and Bob try to adversarial self-test. It turns out that Alice or Bob can always bias the marginals of non-product distributions in quantum settings which we show follows from the insecurity of certain tasks in quantum two-party cryptography. The literature on this area contains many impossibility results.\old{, even when using quantum mechanics. }
Some of the more popular tasks include  
bit-commitment~\cite{mayers1997unconditionally, lo1998quantum, chaillouxbcoptimal}, 
strong coin-flipping~\cite{kitaev2002quantum, chaillouxscfoptimal}, 
die-rolling~\cite{aharon2010quantum, cryptography1020011}, oblivious transfer~\cite{chailloux2010lower, chaillouxsotoptimal, Kundu2022deviceindependent}, and, more generally, secure function evaluation~\cite{osborn2022constant, buhrman2012complete}
(many of the references above point towards their impossibility). 
With this said, we revisit the theme of this paper, stated in the contrapositive.

\medskip
\textit{Theme, restated: The insecurity in quantum cryptography suggests the inability to self-test.}

\medskip 

To illustrate this connection, we consider %
one particular task within two-party cryptography we alluded to earlier, \emph{coin flipping}, where Alice and Bob wish to generate a shared uniformly random bit. 
In other words, they wish to securely sample from the joint distribution $p(ab) = \frac{1}{2} \delta_{a,b}$. 
However, there is a constant lower bound on the security of any quantum coin flipping protocol due to Kitaev~\cite{kitaev2002quantum} indicating that this particular distribution cannot be securely sampled. Can one say something more generally?

Indeed, Kitaev's lower bound states that for \emph{any} quantum protocol that samples from the joint distribution $p(ab)$, we must have  
\begin{equation}\label{eq:kitaev:lowerbound}
    p^*(a) \cdot p^*(b) \geq p(ab) 
\end{equation}
for any fixed $a$ and $b$, where we use the following notation.
\begin{itemize}[leftmargin=*] 
\item $p(ab)$: The probability with which Alice and Bob output $a$ and $b$ (when both follow the protocol honestly).
\item $p^*(a)$: The maximum probability Bob can force Alice to output $a$ (when she follows the protocol honestly).
\item $p^*(b)$: The maximum probability Alice can force Bob to output $b$ (when he follows the protocol honestly).  
\end{itemize} 
Now, suppose one can securely sample from the non-product distribution $p(ab)$. 
Then, for any fixed $\delta > 0$, there exists a protocol such that 
$p(a) + \delta \geq p^*(a)$ and $p(b) + \delta \geq p^*(b)$. 
Combining with Kitaev's bound, we have 
\begin{equation}\label{eq1}
    (p(a) + \delta)(p(b) + \delta) \geq p^*(a) \cdot p^*(b) \geq p(ab)  
\end{equation}
for any $a$ and $b$. 
By taking limits as $\delta \to 0$, we have that $p(a) \cdot p(b) \geq p(ab)$ for all $a$ and $b$ which can only hold for product distributions from our lemma---a contradiction.
Thus, $p(ab)$ cannot be securely sampled, completing the proof of Theorem~\ref{thm2_}.

\medskip 
\paragraph{\textbf{Multiparty setting.}}  
We now briefly discuss the possibility of secure sampling if there are more than two parties. 
Consider the task where there are $n$ parties who wish to sample from the joint distribution $p(a_1 a_2 \ldots a_n)$ where party $i$ outputs $a_i$. 
One may wonder if there is some way to use the extra parties involved to test the devices. 
It turns out that this is also impossible for certain distributions. 
We say that a mulitpartite distribution is non-trivial if there exists a partition such that $p(a_1 a_2 \ldots a_n)$ is non-product across that partition. 
Consider such a non-trivial multipartite distribution and call the partitions $A$ for ``Alice'' and $B$ for ``Bob'' (this suggestive naming convention will make sense shortly).  
If we also set $a$ to be the tuple $(a_i : i \in A)$, and $b$ to be the tuple where $(a_i : i \in B)$, we have effectively reduced the multiparty setting to the two-party setting where we only have Alice and Bob. 
Since $p(ab)$ is non-trivial, then we cannot securely sample from this distribution. 
Here, when Alice or Bob cheats, we suppose that they are not bound by any locality constraints. 
That is, they are allowed to act as a single cheating entity. 
In summary, we cannot securely sample a non-trivial multiparty distribution since there exists one subset of the parties who can (collectively) cheat the rest. 
Thus, if we generalize adversarial self-testing to multiple parties, we see that this is impossible as well if a certain partition and choice of inputs leads to a non-product distribution. 

\medskip 
\paragraph{\textbf{Comparisons to previous work.}}  
The work~\cite{MillerShi}, considers non-local games in a two-phase setting where in the first phase, the two parties cooperate to play the game and in the second phase, they try to learn the other party's output. It shows that even if the other party's input is revealed, their output remains random. 
The multiparty case with dishonest parties has been studied in~\cite{PhysRevLett.131.140201}. 
In this work, the identity of the malicious parties is known in advance. In \cite{DIWCF}, these ideas are applied to improve protocols for the cryptographic task of weak coin flipping.
All these works can be interpreted as positive results as contrasted to our negative result. Indeed, they all have assumptions concerning how the parties trust each other (while we make no assumptions).  
This opens the question of what possible results can be obtained in this adversarial setting if one places assumptions/restrictions on the cheating parties involved.

\medskip 
\paragraph{\textbf{Conclusions.}} 

In this Letter, we proved that one cannot securely sample non-product probability distributions and thus adversarial self-testing of devices producing non-product correlations is also impossible. 
For future work, it would be interesting to see if a form of adversarial self-testing is possible in certain multiparty settings. Perhaps restricting the cheating subsets to only have a small number of parties would circumvent the impossibility.
Another interesting research avenue would be to see how our results change if one were to add restrictions to what Alice and Bob are allowed to do when tampering with the boxes. 
Kitaev's lower bound is in the information-theoretic setting; it may not hold if we impose certain restrictions on Alice and Bob (e.g., Alice and Bob are computationally bounded). 
We believe that adversarial self-testing should be possible in this restricted setting.  
  
\section*{Acknowledgements}  
We are thankful to James Bartusek for helpful discussions.
AB is partially supported by a Bitshares Fellowship. 
ASA acknowledges support from IQIM, an NSF Physics Frontier Center (GBMF-1250002), MURI grant FA9550-18-1-0161 and the U.S. Department of Defense through a QuICS Hartree Fellowship. Part of the work was carried out while ASA was visiting the Simons Institute for the Theory of Computing.
TVH acknowledges support from ParisRegionQCI, supported by Paris Region; FranceQCI, supported by the European Commission, under the Digital Europe program; QSNP: European Union’s Horizon Europe research and innovation program under the project ``Quantum Security Networks Partnership''.
JS is partially supported by Commonwealth Cyber Initiative SWVA grant 467489 and by Virginia Tech's Open Access Subvention Fund.

\bibliographystyle{ieeetr} 
\bibliography{references} 
  
\end{document}